\documentclass{PoS}
\usepackage{amssymb}   
\usepackage{amsmath}   
\usepackage{axodraw}

\input pix.sty

\def\undertilde#1{\mathop{\vtop{\ialign{##\cr$\textstyle{#1}$\cr%
\noalign{\kern1pt\nointerlineskip}\hfil$\mathchar"0365$\hfil\cr}}}}
\def\wideundertilde#1{\mathop{\vtop{\ialign{##\cr$\textstyle{#1}$\cr%
\noalign{\kern1pt\nointerlineskip}\hfil$\mathchar"0367$\hfil\cr}}}}

\renewcommand{\eq}{eq.~}
\renewcommand{\eqs}{eqs.~}

\renewcommand{\fig}{fig.~}

\newcommand{\tinymsbar}{{\overline{\mbox{\tiny\rm{MS}}}}}
\newcommand{\Lambdamsbar}{{\Lambda_\tinymsbar}}
\newcommand{\mD}{m_\rmi{D}}

\newcommand{\CF}{C_\rmii{F}}
\newcommand{\TF}{T_\rmii{F}}
\newcommand{\Nf}{N_{\rm f}}

\newcommand{\Nc}{N_{\rm c}}

\newcommand{\gammaE}{\gamma_\rmii{E}}
\newcommand{\rmO}{{\mathcal{O}}}
\newcommand{\bmu}{\bar\mu}
\newcommand{\CA}{\Nc}
\def\lsi{\raise0.3ex\hbox{$<$\kern-0.75em\raise-1.1ex\hbox{$\sim$}}}
\def\gsi{\raise0.3ex\hbox{$>$\kern-0.75em\raise-1.1ex\hbox{$\sim$}}}

\newcommand{\gsim}{\mathop{\gsi}}

\newcommand{\rmii}[1]{{\mbox{\tiny\rm{#1}}}}

\newcommand{\Tint}[1]{{\hbox{$\sum$}\!\!\!\!\!\!\!\int\,}_{\!\!\!\!\raise-0.9ex\hbox{$\scriptstyle{#1}$}}}
\newcommand{\Tinti}[1]{{{\Sigma}\!\!\!\!\raise0.3ex\hbox{$\int$}_\rmii{${#1}$}}}

\newcommand{\bi}{\begin{itemize}}
\newcommand{\ei}{\end{itemize}}
\newcommand{\hide}[1]{ }

\title{Temporal mesonic correlators at NLO for any quark mass}

\ShortTitle{Temporal mesonic correlators at NLO for any quark mass}

\author{\speaker{Y.\ Burnier}${\,}^a$ and M.\ Laine$^b$\\
        \llap{$^a$}
        Institute of Theoretical Physics, EPFL, 
        CH-1015 Lausanne, Switzerland\\
        \llap{$^b$}
        Institute for Theoretical Physics, AEC, University of Bern, 
        CH-3012 Bern, Switzerland\\
        E-mail: \email{yannis.burnier@epfl.ch}, \email{laine@itp.unibe.ch}}

\abstract{
We present NLO results for thermal imaginary-time correlators in
the vector and scalar channels as a function of the quark mass. The range of
quark masses for which a non-relativistic approximation works in the
temperature range considered is estimated, and charm quarks turn out to be a
borderline case.  Comparing with simulation data from fine lattices, we find
good agreement in the vector channel but a substantial discrepancy in the
scalar one. An explanation for the discrepancy is suggested in terms of
physics of the quark-antiquark threshold region. Perturbative predictions 
for the bottom scalar spectral function around the threshold are also 
briefly reviewed.
}

\FullConference{31st International Symposium on Lattice Field Theory 
                LATTICE 2013\\
		July 29 - August 3, 2013\\
		Mainz, Germany}

\begin{document}

%
\section{Motivation}

Massive quarks have long been considered as excellent probes for 
the physics of a quark-gluon plasma. Due to their heavy mass they could
experience changes at finite temperature that are both theoretically
tractable and experimentally identifiable. The fortunate existence
of both charm and bottom quarks in an appropriate mass range implies
that the quark mass can be considered a tunable parameter  
and that theoretical predictions may be interpolated or 
extrapolated as a function of the quark mass. The fates
of ``open'' heavy-flavour D and B mesons on one hand, and ``bound''
charmonium and bottomonium systems on the other, capture a rich
spectrum of interesting physics phenomena. Heavy quarks are also 
relatively easy to simulate on the lattice, even though care needs
to be taken in view of possible discretization artefacts. 

The present study is related to lattice 
measurements of two-point correlators 
of heavy scalar densities and vector currents at finite temperature. 
Ultimately, 
the goal is to use imaginary-time correlators measured on the lattice
in order to constrain the corresponding spectral functions; the latter, 
in turn, describe open heavy flavour physics through a transport peak
at small frequency, and quarkonium physics through a threshold region
at large frequency. In the present investigation the main focus is on 
imaginary-time correlators, which we have recently computed up to 
next-to-leading order (NLO) as a function of the quark 
mass~\cite{GVtau,GStau} and compared with quenched data from 
fine lattices~\cite{ding2}.

%
\section{Charm quark imaginary-time correlators}

\begin{figure}[t]


\centerline{%
 \epsfysize=7.8cm\epsfbox{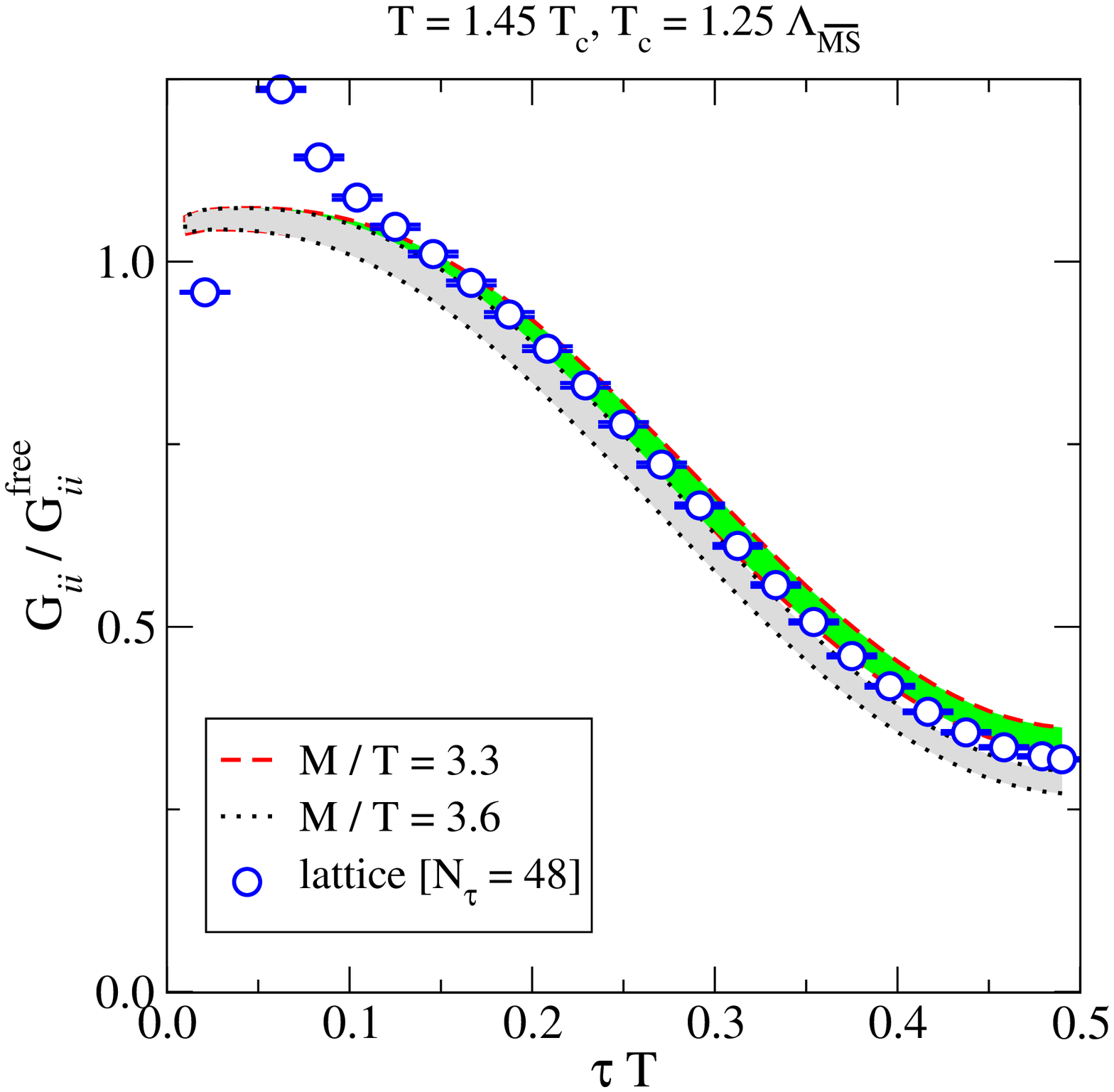}%
~~~\epsfysize=7.8cm\epsfbox{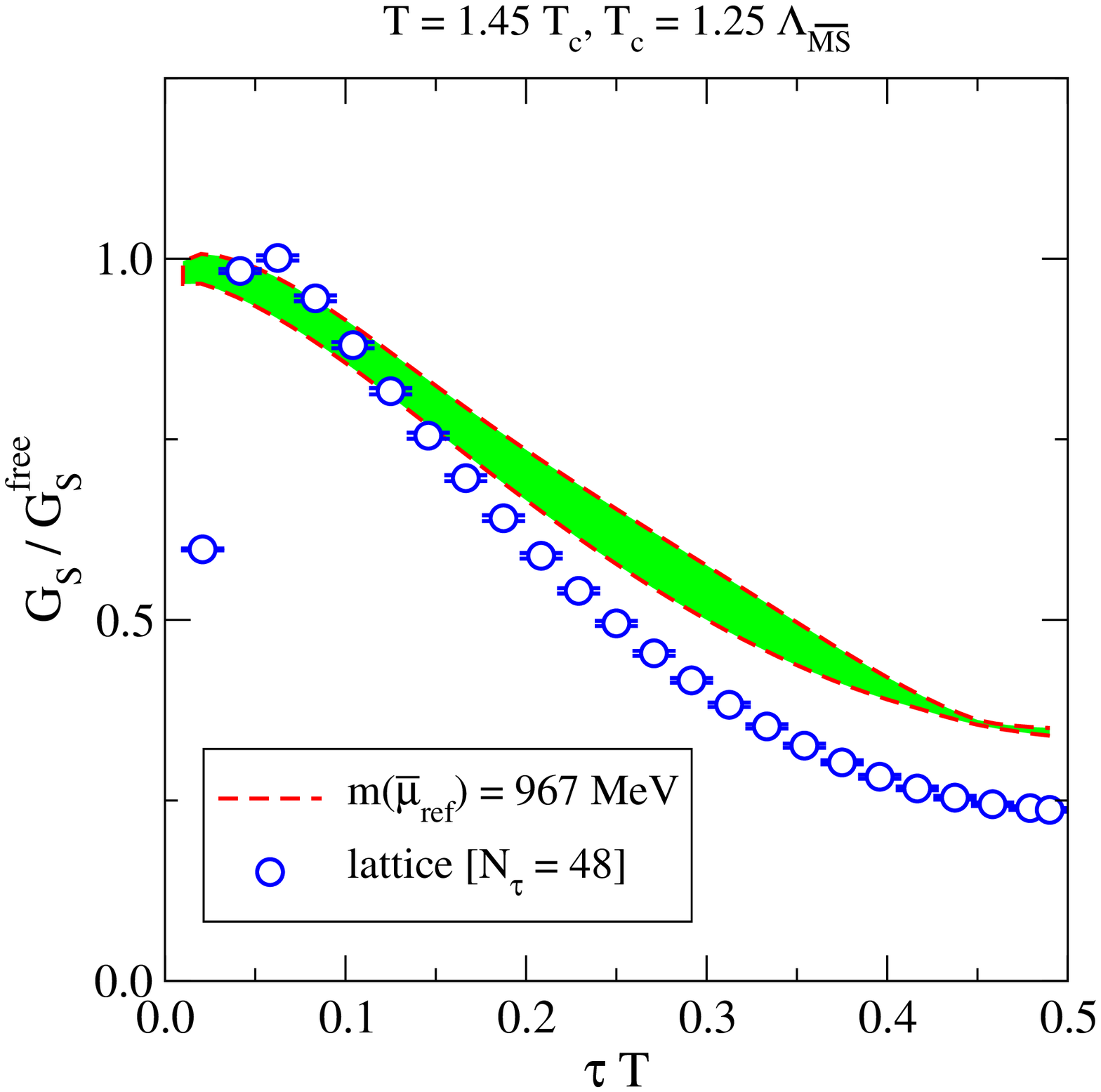}
}

\caption[a]{
Left: The spatial part of the vector correlator, \eq\nr{GV_def}, 
normalized to \eq\nr{GV_norm}, 
compared with lattice data from ref.~\cite{ding2}. 
The pole masses $M/T = 3.3, 3.6$ are chosen from a perturbative
estimate and from an optimal agreement of 
quark number susceptibility, respectively~\cite{GVtau}. 
Right: The scalar correlator, \eq\nr{GS_def}, normalized to \eq\nr{GS_norm}, 
compared with lattice data from ref.~\cite{ding2}. The $\msbar$ mass
$m(\bmu^{ }_\rmi{ref}) = 967$~MeV, 
with $\bmu^{ }_\rmi{ref} = 2$~GeV~\cite{pdg},  
corresponds to the value
$m_c(m_c) = 1.094(1)$~GeV cited in ref.~\cite{ding2}.
(In the scalar channel the pole mass 
scheme shows questionable convergence already at NLO~\cite{GStau}.)
}

\la{fig:optimal}
\end{figure}

The vector and scalar correlators are defined in continuum as 
\ba
 G^{ }_{ii}(\tau)
 & \equiv & 
 \sum_{i = 1}^{3}
 \int_\vec{x} 
 \Bigl\langle 
 (\bar\psi \gamma_i \psi) (\tau,\vec{x}) 
 \;
 (\bar\psi \gamma_i \psi) (0,\vec{0})
 \Bigr\rangle^{ }_T
 \la{GV_def} \;, \\
 G^{ }_\rmii{S}(\tau)
 & \equiv & 
 M_\rmii{B}^2 \int_\vec{x} 
 \Bigl\langle 
 (\bar\psi \psi) (\tau,\vec{x}) 
 \;
 (\bar\psi \psi) (0,\vec{0})
 \Bigr\rangle^{ }_T
 \;, \la{GS_def} 
\ea
where $M^{ }_\rmii{B}$ is the bare quark mass
and $ 0 < \tau < {1} / {T} $. 
As is clear from 
the definitions, the scalar correlator is more sensitive
to the quark mass than the vector one. (Without the 
bare quark mass in the definition, 
the scalar correlator would not be renormalizable
even at NLO, and it would lose its connection to the QCD Lagrangian.)

Physically, the charm quark vector correlator 
is related to an in-medium $J/\psi$ contribution to 
the thermal dilepton production rate, 
as well as to the charm quark diffusion 
coefficient and kinetic equilibration rate. 
The scalar density operator represents the quark
contribution to the trace of the energy-momentum tensor and 
is hence related to the bulk viscosity 
of the quark-gluon plasma
as well as to the charm
quark chemical equilibration rate. In addition, the scalar
spectral function around threshold is believed to describe $P$-wave
charmonium states. 

The lattice and NLO results for the two correlators 
are compared in \fig\ref{fig:optimal}. 
The results are normalized to massless ``free'' correlators, 
\ba
 G_{ii}^\rmi{free}(\tau) & \equiv &   
 2 \Nc T^3 \biggl[ 
  \pi \, (1-2\tau T) \,
  \frac{1 + \cos^2(2\pi\tau T)}{\sin^3(2\pi \tau T)}
 + 
 \frac{2 \cos(2\pi \tau T)}{\sin^2(2\pi \tau T)}
  + \fr16 \biggr]
 \;, \la{GV_norm}  \\
  G_\rmii{S}^\rmi{free}(\tau) & \equiv &   
 \Nc T^3  
  m^2_{\tau}
 \biggl[ 
  \pi \, (1-2\tau T) \,
  \frac{1 + \cos^2(2\pi\tau T)}{\sin^3(2\pi \tau T)}
 + 
 \frac{2 \cos(2\pi \tau T)}{\sin^2(2\pi \tau T)}
  \biggr]
 \;, \la{GS_norm} 
\ea
where we have defined
$
  m^2_{\tau}  \equiv 
  m^2(\bmu_\rmi{ref})
 \bigl\{ 
        {\ln\bigl[ \frac{\bmu_\rmi{ref}}{\Lambdamsbar} \bigr] } / 
        {\ln\bigl[ 
 \frac{\beta e^{\frac{1}{12} - \gammaE} }
   {\tau(\beta-\tau )\Lambdamsbar} \bigr]} 
 \bigr\} ^{\frac{18 \CF}{11\CA - 4 \TF}}
$,
$\bmu^{ }_\rmi{ref} \equiv $~2 GeV~\cite{pdg}, 
$\beta \equiv 1/T$, 
$\CF \equiv (\Nc^2-1)/(2\Nc)$, 
and $\TF \equiv \Nf/2$.
Moreover $m(\bmu)$ is the $\msbar$ scheme quark mass.

It can be observed from \fig\ref{fig:optimal} that in the vector
channel the results agree well (apart from discretization artefacts
at small $\tau$). 
In contrast, in the scalar 
channel a clear discrepancy is visible. In the following, 
we concentrate on understanding what is going on in the scalar correlator. 

%
\section{Validity of the non-relativistic approximation}

\begin{figure}[t]


\centerline{%
 \epsfysize=7.8cm\epsfbox{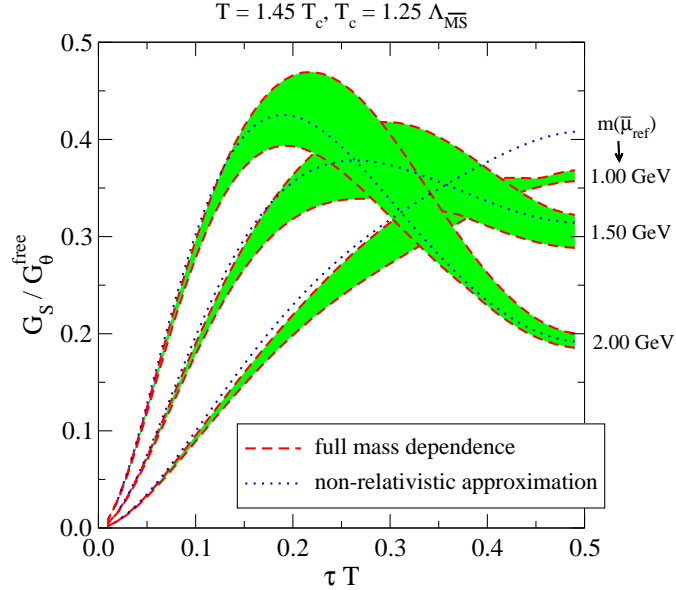}%
}

\caption[a]{
Comparison of NLO scalar correlators with full mass
dependence (bands), against results based on a non-relativistic approximation
of the NLO spectral function at large $\omega$~\cite{nlo} 
together with a constant contribution from the transport peak 
at small $\omega$~\cite{GStau} (dotted lines), normalized
to \eq\nr{Gtheta_norm}. 
The non-relativistic approximation
is accurate for $m(\bmu^{ }_\rmi{ref}) \gsim 1.5$~GeV
at this temperature, which is above the physical value
$m_c(\bmu^{ }_\rmi{ref}) = 1.275(25)$~GeV~\cite{pdg}. However, 
even for $m(\bmu^{ }_\rmi{ref}) = 1.0$~GeV, its breakdown is not catastrophic.
}

\la{fig:nonrel}
\end{figure}

In order to investigate possible reasons for the discrepancy
in \fig\ref{fig:optimal}(right), it is 
useful to view the imaginary-time correlator as originating from an 
underlying spectral function $\rho^{ }_\rmii{S}$,
\be
 G^\rmi{ }_\rmii{S}(\tau) =
 \int_0^\infty
 \frac{{\rm d}\omega}{\pi} \rho^\rmi{ }_\rmii{S} (\omega)
 \frac{\cosh \bigl[ \bigl(\frac{\beta}{2} - \tau\bigr)\omega\bigr]}
 {\sinh\bigl(\frac{\beta \omega}{2}\bigr)} 
 \;. \la{relation}
\ee
The spectral function represents the cut (imaginary part) of a 
two-point correlator in momentum space. Surprisingly, spectral
functions are not particularly well studied in the presence of a quark
mass: even in vacuum, the result is known analytically only up 
to NLO (cf.\ ref.~\cite{old} and references therein; numerical estimates
exist also at higher orders, cf.\ ref.~\cite{nnlo}). 
For $m\neq 0$ NLO
thermal corrections to $\rho^{ }_\rmii{S}$
have been computed only for the
``non-relativistic'' regime $m \gg \pi T$~\cite{nlo}. (This is 
peculiar since the generalization of massless 
zero-momentum NLO spectral functions to a finite mass
should be less complicated than 
to a finite momentum~\cite{dilepton}.)

It is useful to start by asking how well the non-relativistic results of 
ref.~\cite{nlo} compare with the numerically determined imaginary-time
correlators computed in ref.~\cite{GStau}, where no approximation was
made with respect to the quark mass. For this purpose, 
the contribution of the transport peak, which was not addressed in 
ref.~\cite{nlo}, needs to be added. Within NLO perturbation theory 
the transport peak yields an exactly $\tau$-independent contribution, 
given in \eqs(4.8), (4.10), (4.13) of ref.~\cite{GStau}. 
We sum this to the contribution from the spectral function, 
and normalize the results to a purely gluonic scalar correlator, 
{\em viz.}
\ba
 \frac{ G_{\theta}^\rmi{free}(\tau) }{2  \CA \CF T^5} 
 & \equiv &   
  (8 \pi c_\theta g_\tau^2 )^2 
 \biggl[ 
  \pi \, (1-2\tau T) \, 
  \frac{2\cos(2\pi\tau T)  + \cos^3(2\pi\tau T)}{\sin^5(2\pi \tau T)}
 + 
 \frac{1 + 2 \cos^2(2\pi \tau T)}{\sin^4(2\pi \tau T)}
  \biggr]
 \;, \hspace*{5mm} \la{Gtheta_norm} 
\ea
where
$  g_\tau^2 \equiv  
        {24 \pi^2 } / \{ 
        {(11 \CA - 4 \TF)
       \ln\bigl[ \frac{\beta e^{\frac{14}{33} - \gammaE}}
      {\tau(\beta-\tau )\Lambdamsbar} \bigr]} \}
$,  
$
  c^{ }_\theta \equiv   - {b^{ }_0} / {2} - {b^{ }_1 g_\tau^2} / {4} 
$, 
and $b_0, b_1$ are coefficients of the QCD $\beta_g$-function. 
The justification for this normalization is that it is mass-independent 
and conveniently magnifies the interesting large-$\tau$ regime.  

The result of the comparison is shown in \fig\ref{fig:nonrel}.
We observe that the non-relativistic approximation is 
accurate for $m(\bmu^{ }_\rmi{ref}) = 2$~GeV, whereas 
for $m(\bmu^{ }_\rmi{ref}) = 1$~GeV
(which is 
close to $m(\bmu^{ }_\rmi{ref}) = 967$~MeV simulated in ref.~\cite{ding2})
a discrepancy is visible. The physical case, with 
$m_c(\bmu^{ }_\rmi{ref}) = 1.275(25)$~GeV~\cite{pdg}, lies
in between, however surely not deep in the non-relativistic regime. 
This is interesting in its own right, because if 
the charm quarks are not really exponentially suppressed,
then the question of their partial chemical equilibration 
may be raised~\cite{chem}. 

%
\section{Explaining the discrepancy in the scalar channel}

\begin{figure}[t]


\centerline{%
 \epsfysize=7.8cm\epsfbox{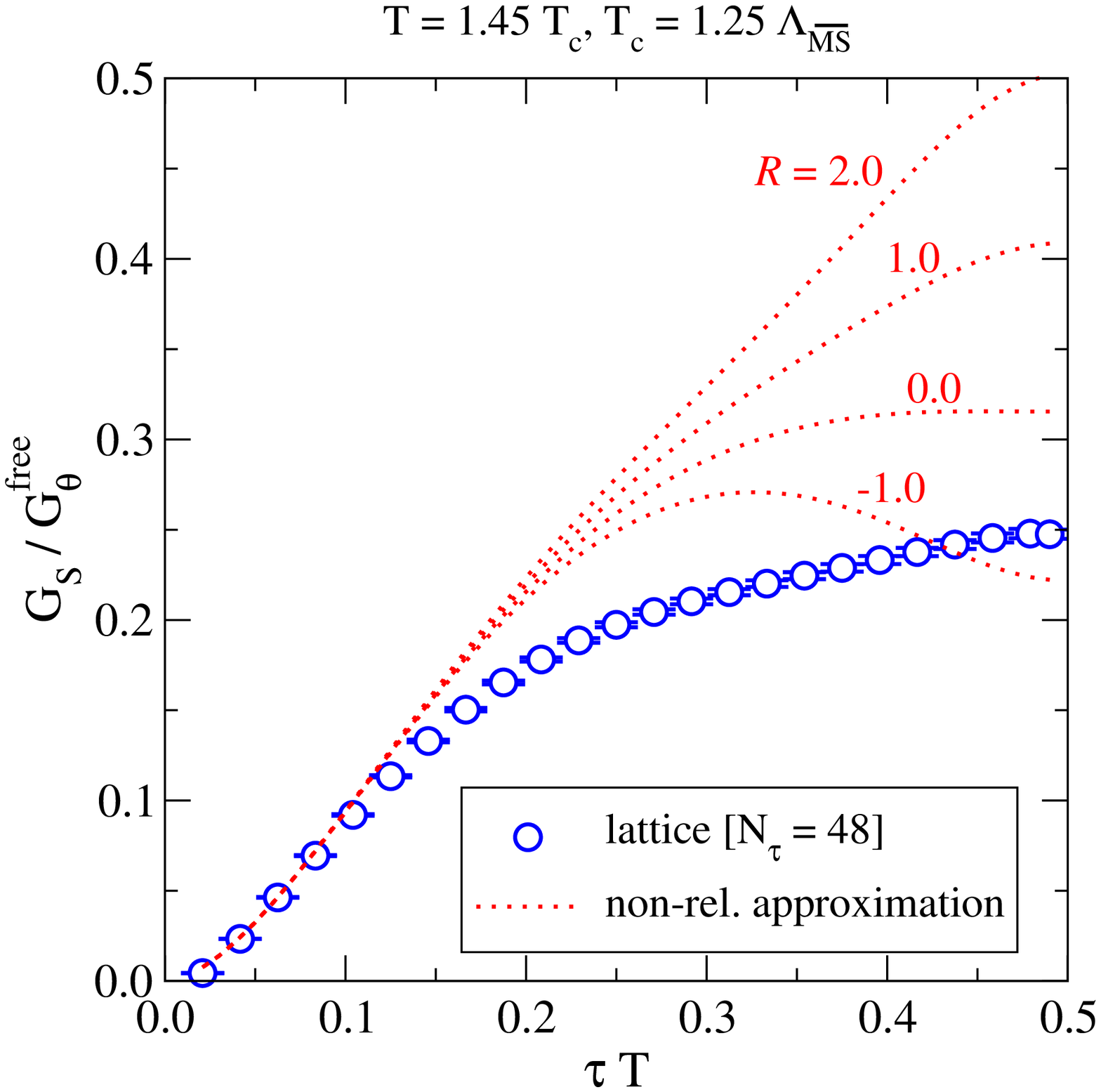}%
~~~\epsfysize=7.8cm\epsfbox{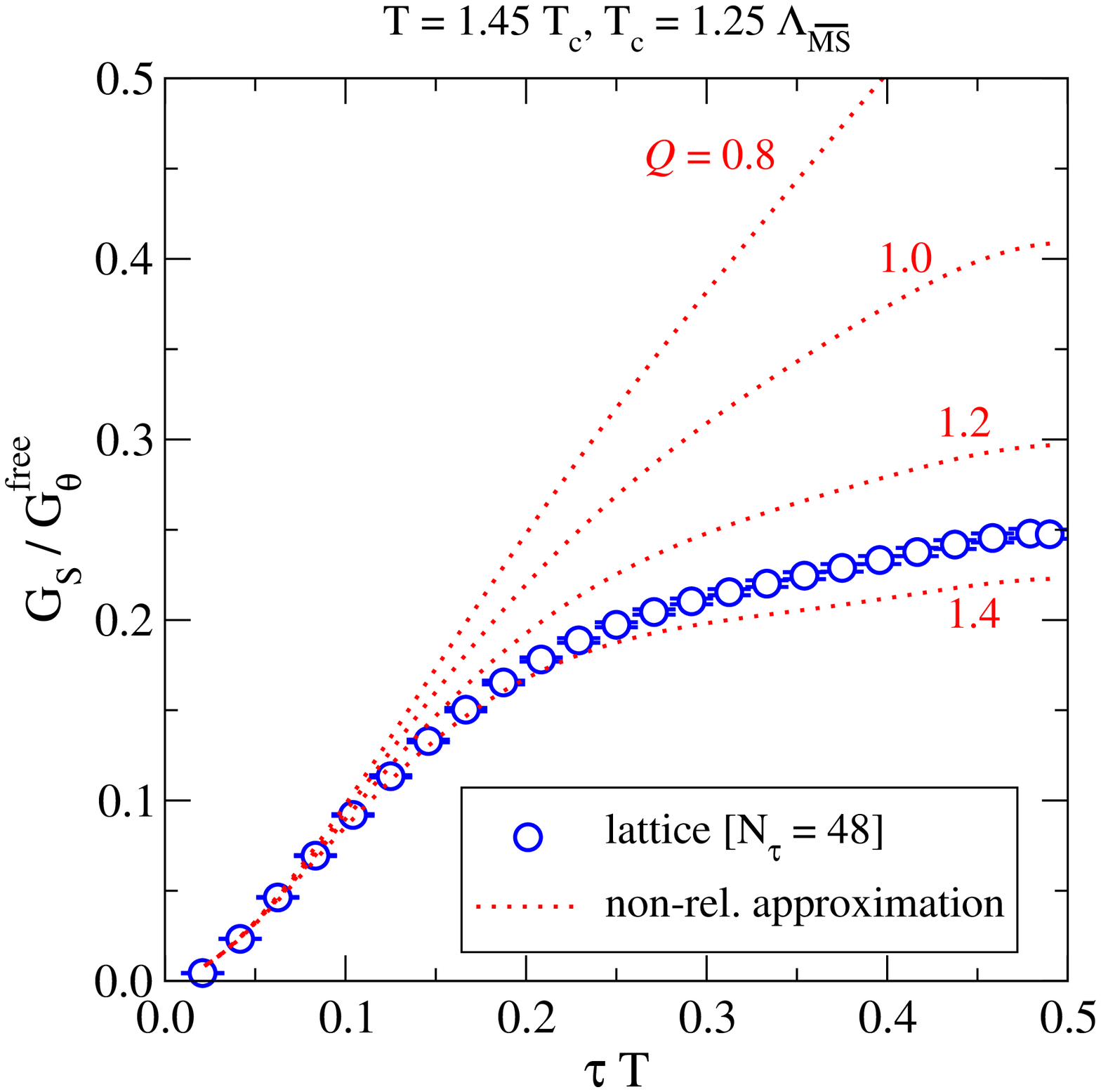}
}

\caption[a]{
Left: Modification of the scalar correlator if the 
contribution from the transport peak is multiplied by 
a factor ${R}$, cf.\ \eq\nr{calR}. 
It is seen that even a large deviation from $R = 1$ does not help.
Right: Modification of the scalar correlator
if the threshold location is shifted left or right by
a multiplicative factor $Q$, cf.\ \eq\nr{calQ}. 
A substantial improvement can be observed. Both plots are 
based on a non-relativistic approximation in the regime 
$\omega > m (\bmu_\rmi{ref}) = 967$~MeV, omitting terms suppressed
by $e^{-m (\bmu_\rmii{ref}) /T}$.
}

\la{fig:discrepancy}
\end{figure}

Moving on, we have carried out two tests in order to probe 
the origins of the discrepancy seen in \fig\ref{fig:optimal}(right). 
The first test concerns the contribution of the transport peak.
As mentioned, this yields a  
constant contribution 
within NLO perturbation theory. We have tested how changing
the amplitude of the constant by a factor ${R}$ changes
the result: 
\be
 \left. G^\rmii{LO+NLO}_\rmii{S} \right|_\rmi{const.}
 \to  
 \left. G^\rmii{LO+NLO}_\rmii{S} \right|_\rmi{const.}
  \times {R} 
 \;.  \la{calR}
\ee
The result is shown in \fig\ref{fig:discrepancy}(left), and we find
no substantial improvement. 

\begin{figure}[t]


\centerline{%
 \epsfysize=7.5cm\epsfbox{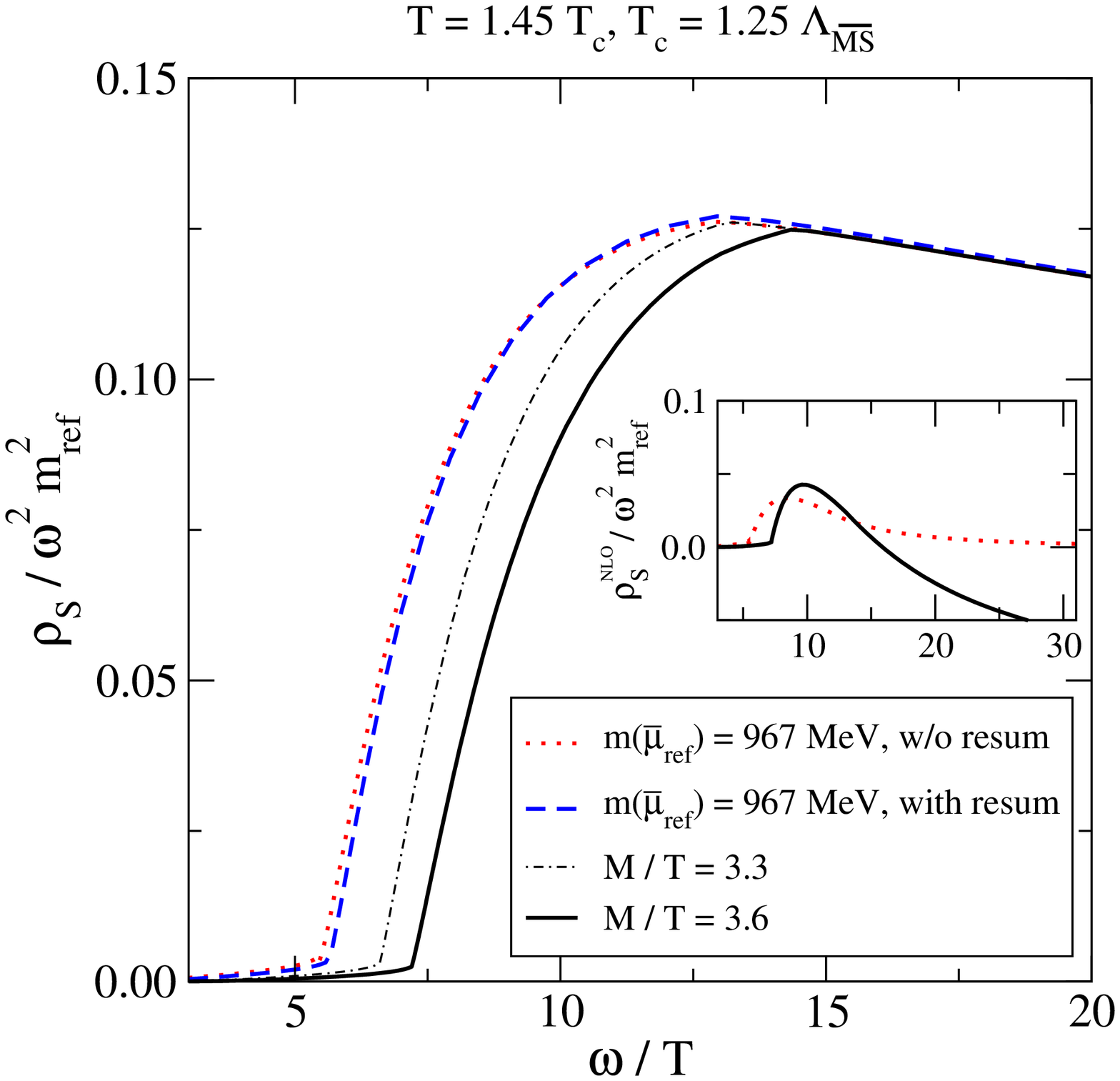}%
~~~\epsfysize=7.5cm\epsfbox{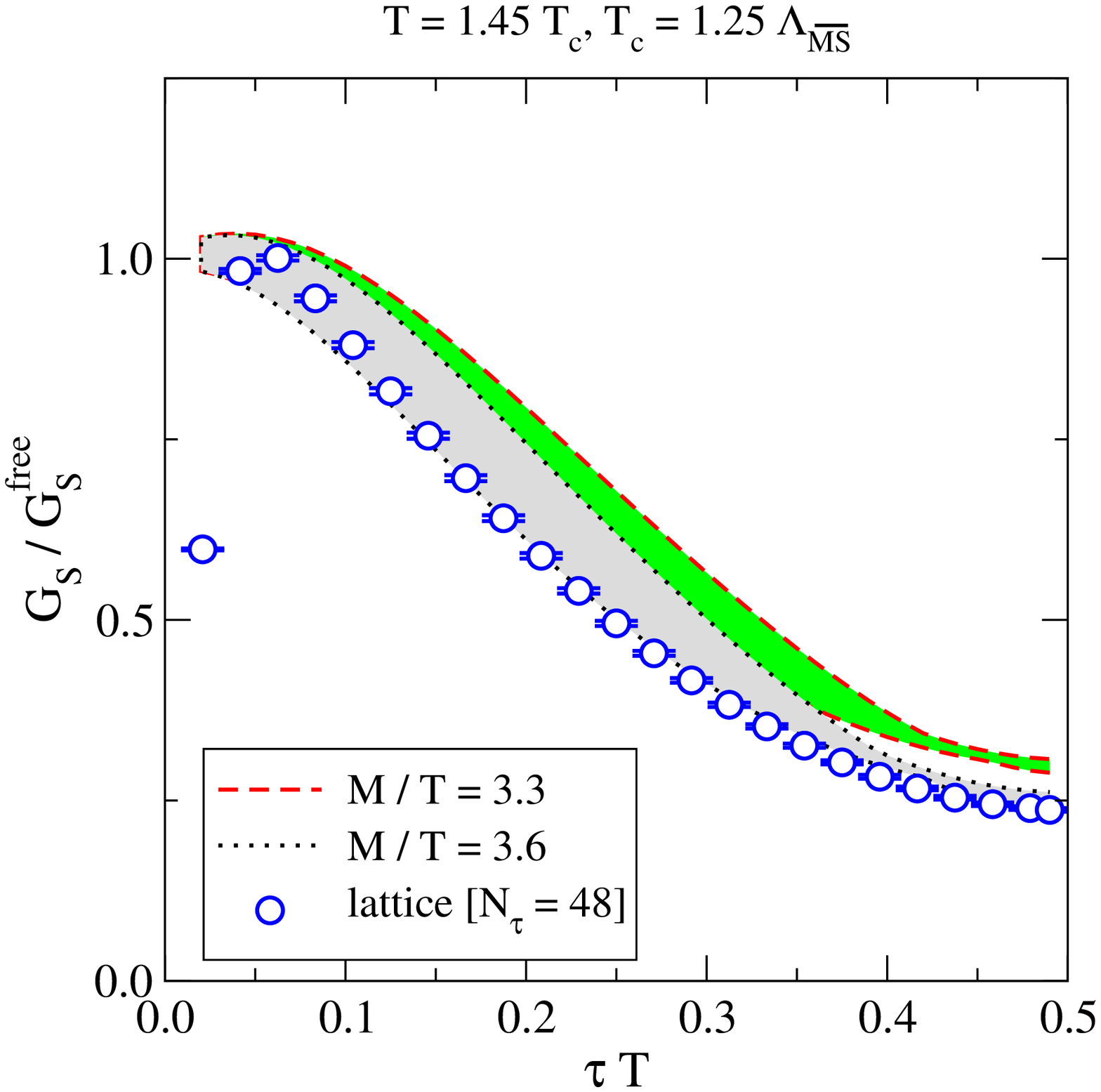}
}

\caption[a]{
Left: The LO+NLO scalar channel spectral function around the threshold, 
with and without mass resummation in the $\msbar$ scheme 
(cf.\ \eq\nr{mass_resum}), and with mass resummation 
in the pole mass scheme. The inset shows the NLO parts
and illustrates the non-convergence of
the pole mass result at large $\omega$.  
Right: The imaginary-time correlator, 
from a rescaled pole mass result at $\omega < 4 M$
and $\msbar$ result at $\omega > 4 M$, compared with
lattice data~\cite{ding2}. The remaining discrepancy is probably due to 
the non-relativistic approximation. 
}

\la{fig:rho}
\end{figure}

The second test concerns the threshold region. 
Thermal corrections modify the threshold location by a well-known 
NLO correction~\cite{dhr}, 
\be
 m^2 \to m^2 + {g^2 T^2 \CF} / {6}
 \;, \la{mass_resum}
\ee 
which tends to move the threshold to larger frequencies. For a large
quark mass, however, the effect is small: 
$
 \delta \omega_\rmi{threshold} = 
 {g^2 T^2 \CF}/ { (12 m) }
$.
There is an effect of opposite sign originating
from a Debye-screening induced correction to a heavy quark mass,
$
 \delta \omega_\rmi{threshold} = - {g^2\CF \mD}/ {(4\pi)}
$~\cite{rdp}. However, apart from these
thermal corrections, there is also 
an important zero-temperature effect:
whereas accounting for the small-$\tau$ behaviour of $G^{ }_\rmii{S}$
requires the use of the $\msbar$ scheme or a 
similar running mass~\cite{GStau}, 
it is known that 
threshold features are better described by a pole-type mass
(cf.\ e.g.\ ref.~\cite{mb}). 
Like in \fig\ref{fig:optimal}
we denote the pole mass by $M$.

In order to probe these effects, we have considered a shift of 
the threshold location by multiplying, in effects originating from 
quark propagators, the quark mass by a factor $Q$. However
the overall multiplicative factor $m^2$ originating from \eq\nr{GS_def}
is left unchanged. With a particular scale choice~\cite{GStau}, 
the location of the zero-temperature threshold 
is thus given by the solution of
\be
 \omega_\rmi{threshold} = 2 
 m(\bmu = \omega_\rmi{threshold}  \, e^{-17/12}) \times {Q} 
 \;. \la{calQ}
\ee
The result is shown in \fig\ref{fig:discrepancy}(right);
the discrepancy is considerably reduced for $Q > 1$. 
If we recall
from \fig\ref{fig:nonrel} that the non-relativistic approximation 
overestimates the true answer at these quark masses, the optimal 
value might be $Q \simeq 1.2$ or so. Remarkably, this is quite 
close to the ratio of the pole and $\msbar$ masses for the parameter
values used in ref.~\cite{ding2}, 
\be
 \frac{ M } {m_c (m_c) } = 
 1  + \frac{4 g^2 (m_c) \CF}{(4\pi)^2} 
 + \rmO(g^4)
 \approx \frac{1.3~\mbox{GeV}}{1.1~\mbox{GeV}} \approx 1.2
 \;. 
\ee 

As a crosscheck, we have employed the same pole masses 
$M/T \approx 3.3, 3.6$ as in \fig\ref{fig:optimal}(left) 
for treating the scalar channel threshold region. 
More concretely, we have considered the NLO spectral function as given 
in ref.~\cite{nlo}, which was indeed in the pole mass scheme,
as well as the corresponding $\msbar$ scheme one, with or without
thermal mass resummation.\footnote{%
 The $\msbar$ scheme result is obtained from the expressions 
 of ref.~\cite{nlo} by setting $M\to m(\bmu)$ and
 $\delta \to  - \ln[{\bmu^2}/{m^2(\bmu)}] - 4/3$. 
 Thermal mass resummation can be removed by changing 
 $+4 k^2 \to -2 k^2$ in \eq(C.11).} 
The pole mass result is not reliable at large $\omega$, 
because the NLO correction overtakes the LO term 
and the perturbative series breaks 
down, cf.\ the inset in \fig\ref{fig:rho}(left).
We normalize the pole mass result 
such that it agrees with the unresummed $\msbar$ 
result at $\omega = 4 M$;
above this, the $\msbar$ result is used, 
cf.\ \fig\ref{fig:rho}(left). 
The resulting imaginary-time
correlators are illustrated in \fig\ref{fig:rho}(right), together
with a comparison with lattice data. 
The agreement is much better than in \fig\ref{fig:optimal}(right), 
and remarkably good considering that there are errors related
to the non-relativistic approximation as visible in 
\fig\ref{fig:nonrel}.

%
\section{Scalar channel spectral function in the bottom quark case}

In the bottom quark case there is no doubt about the validity 
of the non-relativistic approximation. This permits the use of 
special effective theories, such as Heavy Quark Effective Theory
for addressing the transport region~\cite{eucl} 
and Non-Relativistic QCD for addressing 
the threshold region~\cite{nrqcd2}.
One issue of phenomenological controversy is that whereas there is 
certainly no resonance peak in the scalar spectral function
for $M < 2$~GeV, for the bottom case $M \simeq 4.5$~GeV
a small $S$-channel contribution 
has been suggested to appear in the dominantly
$P$-channel scalar correlator~\cite{peskin}. This induces a peak to 
the corresponding spectral function. It will be interesting to see
whether a peak can be resolved from data~\cite{peak,new}
with refined spectral analysis tools~\cite{br}. 

%
\section{Conclusions}

The study of charm quark correlators in hot QCD may soon
enter a mature phase. On the lattice side a continuum limit
remains to be taken, but the lattices used are already in a scaling
regime, at least in the quenched case. On the continuum side, full NLO
spectral functions need to be computed, going beyond the present 
non-relativistic approximation. Once these steps have been taken, 
the short-$\tau$ regimes of the two sides should match, and a non-divergent
difference from the large-$\tau$ regime could be subjected to a spectral
analysis of genuine non-perturbative effects. 

This work was partly supported SNF
under the grants 200021-140234 and PZ00P2-142524. 

%

\end{document}